\documentclass{aa}  
\usepackage{graphicx}
\usepackage{txfonts}
\usepackage{color}
\usepackage{xcolor}
\usepackage{eufrak}
\usepackage{graphicx}
\usepackage[mathscr]{euscript}
\definecolor{v}{rgb}{0.54, 0.17, 0.89} 
\usepackage{hyperref}
\usepackage[utf8]{inputenc}
\usepackage{amssymb}
\usepackage{newunicodechar}
\usepackage{hyperref}
\graphicspath{{figures/}}

\begin{document}

   \title{First black hole mass estimation for the quadruple lensed system WGD2038-4008}

   \author{A. Melo
          \inst{1}
       \and   V. Motta\inst{1}
      \and    N. Godoy\inst{1,2}
      \and    J. Mejia-Restrepo\inst{3}
      \and   Roberto J. Assef\inst{4}
      \and    E. Mediavilla\inst{5,6}
      \and    E. Falco\inst{7}
      \and    F. Ávila-Vera\inst{1}
      \and    R. Jerez\inst{1}
          }
\institute{Instituto de F\'isisca y Astronom\'ia, Facultad de Ciencias, Universidad de Valpara\'iso,
Av. Gran Bretaña 1111, Valparaíso, Chile.
\email{alejandra.melo@postgrado.uv.cl}
\and
Núcleo Milenio de Formación Planetaria – NPF, Universidad de Valparaíso, Av. Gran Bretaña 1111, Valparaíso, Chile.
\and
European Southern Observatory, Alonso de C\'ordova 3107, Vitacura, Santiago, Chile.
\and
Núcleo de Astronomía de la Facultad de Ingeniería y Ciencias, Universidad Diego Portales, Av. Ejército Libertador
441, Santiago, 8320000, Chile.
\and
Instituto de Astrofísica de Canarias, Vía Láctea S/N, La Laguna 38200, Tenerife, Spain.
\and
Departamento de Astrofísica, Universidad de la Laguna, La Laguna 38200, Tenerife, Spain.
\and
Harvard-Smithsonian Center for Astrophysics, 60 Garden St., Cambridge, MA 02138, USA.
\\}

\date{September, 2021}

 
  \abstract
   {
   The quadruple lensed system WGD2038-4008 ($\rm z_s=0.777 \pm 0.001$) was recently discovered with the help of new techniques and observations. Even though black hole mass has been estimated for lensed quasars, it has been calculated mostly for one broad emission line of one image, but the images could be affected by microlensing, affecting the results.   
   }
   {
 We present black hole mass (M$_{BH}$) estimations for images A and B of WGD2038-4008 using the three most prominent broad emission lines (H$\alpha$, H$\beta$ and Mg{\footnotesize{II}}) obtained in one single-epoch spectra. This is the first time the mass is estimated in a lensed quasar in two images, allowing us to disentangle the effects of microlensing. The high S/N of our spectra allows us to get reliable results and compare with the existing data in the literature.
   }
   {
   We used the X-shooter instrument mounted in the Very Large Telescope (VLT) at Paranal Observatory, to observe this system taking advantage of its wide spectral range (UVB, VIS and NIR arm). The sky emission correction was made using Principal Component Analysis (PCA) as the nodding was small compared to the image separation. We compare the lines profiles to identify the microlensing in the broad line region (BLR) and corrected each spectra by the image magification. Using the flux ratio between the continuum and the core of the emission lines we analyzed if microlensing was present in the continuum source.}
   {
  We obtained M$_{BH}$ using the single-epoch method with the H$\alpha$ and H$\beta$ emission lines from the monochromatic luminosity and the velocity width. The luminosity at 3000 \AA\, was obtained using the Spectral Energy Distribution (SED) of image A while the luminosity at 5100 \AA\,  was estimated directly from the spectra. The average M$_{BH}$ between the images obtained was $\rm log_{10}$(M$_{BH}/M_{\odot}$) = 8.27 $\pm$ 1.05, 8.25 $\pm$ 0.32 and 8.59 $\pm$ 0.35 for Mg{\footnotesize{II}}, H$\beta$ and H$\alpha$ respectively. 
  We find Eddington ratios similar to those measured in the literature for unlensed low-luminosity quasars.
  Microlensing of -0.16 $\pm$ 0.06 mag. in the continuum was found but the induced error in the $M_{BH}$ is minor compared to the one associated to the macromodel magnification. We also obtained the accretion disk size using the M$_{BH}$ for the three emission lines, obtaining an average value of $\rm log_{10}(r_{s}/cm) = 15.3 \pm 0.63$, which is in agreement with theoretical estimates.
   }
   {}

               
 \maketitle
%
\section{Introduction}

The number of lensed quasars discovered is consistently growing thanks to the help of new identification techniques and observations (\citealt{Agnello2018,2019Kronemartins,2020lemon}). Here, we study one of these recently identified lenses, $\rm WGD2038$-$4008$. This system is a quadruple lensed quasar discovered in 2017 using a combination of Wide-field Infrared Survey Explorer ($WISE$, \citealt{2010Wright}) and $Gaia$ (\citealt{2016Gaiacollab}) over the Dark Energy Survey ($DES$, \citealt{2016DES}) footprint with a source and deflector redshift of $\rm z_s=0.777 \pm 0.001$ and $\rm z_l=0.230 \pm 0.002$ respectively (\citealt{Agnello2018}). The deflector is a red galaxy with a compact bulge and a bright halo while the source has an extended quasar host galaxy (\citealt{Agnello2018}). It has been observed using the Hubble Space Telescope ($HST$) obtaining a lens model for this system using LENSTRONOMY (\citealt{shajib2019}). Spatially resolved narrow-line fluxes ([OIII] in \citealt{Nierenberg2020}) are also available. The lensing galaxy has  been studied in \citealt{Buckley-Geer2020} to measure its velocity dispersion and to identify the line-of-sight galaxies that need to be included in the lens model. Even though gravitational lensed quasars are a powerful tool to study the inner structure of active galactic nuclei (AGNs; \citealt{2007pooley,2008Anguita,2008pointdexter,2010dai,2010morgan,2010treu,2014jimenezvicente,Motta2017}), no such study has been conducted so far for $\rm WGD2038$-$4008$.\\
One of the difficulties of working with lensed quasars is that microlensing could affect different regions of the broad emission lines (BELs) in the spectra of one (or more) image of the system (\citealt{2011mediavilla,Motta2012,Guerras2013,2018fian,Rojas2020}). Microlensing can affect the observed flux of the accretion disk and the BELs, as well as the shape of the BEL, ultimately adding uncertainty to the single-epoch black hole mass (M$_{BH}$) estimation.\\
Precise measurement of the M$_{BH}$ is key in the understanding of the coevolution between the supermassive black hole (SMBH) growth and their host galaxy (see \citealt{2005ferrarese,2013kormendyandho}). In particular, physical parameters of the SMBH seem to correlate well with the luminosity (\citealt{2003Marconi}) and velocity dispersion (\citealt{2000Ferrarese,2002Tremaine}) of the host galaxy.\\
The single-epoch method (SE) is one of the most widely used technique to measure M$_{BH}$ in AGNs, it relates the continuum luminosity of the quasar at a particular wavelength with the size of the broad line region (BLR) (see \citealt{2004ApJvestergaard,2012shenliu,2016MNRAS.460..187M}). Typically, the SE masses for low redshift quasars ($z < 0.7$) are estimated in the optical using H$\alpha$ and H$\beta$ BELs and the continuum luminosity at 5100\AA. However, the Balmer lines are shifted into the infrared at higher redshifts, thus most of the estimations have been measured in the UV wavelength using Mg{\footnotesize{II}} and CIV BELs. Over the last decade, SE method has been used to obtain M$_{BH}$ in lensed AGNs (\citealt{2006Peng,2010greene,2011assef,2012Sluse,2018mediavilla}), but most of them from the Mg{\footnotesize{II}} and CIV BELs, and none of them using different emission lines observed simultaneously.\\
In this paper we present high signal-to-noise ratio (S/N) single-epoch spectra for the quadruple lensed system $\rm WGD2038$-$4008$ to obtain the M$_{BH}$ for three emission lines (H$\alpha$, H$\beta$ and Mg{\footnotesize{II}}) for two of the images. We also study microlensing in the emission lines and in the continuum and finally we obtain the velocity dispersion of the lensing galaxy.\\ 
The paper is organized as follows. In Section~\ref{sec:obs} we present the data along with the reduction and the extraction of the spectra for each component. Section~\ref{sec:Method} shows the description of the method used for the estimation of the M$_{BH}$, microlensing analysis and velocity dispersion of the lensing galaxy. We present our results in Section~\ref{sec:results} comparing with previous studies of different lensed quasars and finally our conclusions are presented in Section~\ref{sec:conclusion}. We assume a $\Lambda$CDM cosmology with: $\Omega_{\Lambda}=0.7$, $\Omega_{M}=0.3$ and $H_{O}=70~ \rm km s^{-1}~Mpc^{-1}$.

\section{Observation and data reduction\label{sec:obs}}

\subsection{Observational strategy}

\begin{figure}
 \centering
  \includegraphics[width=8.0cm]{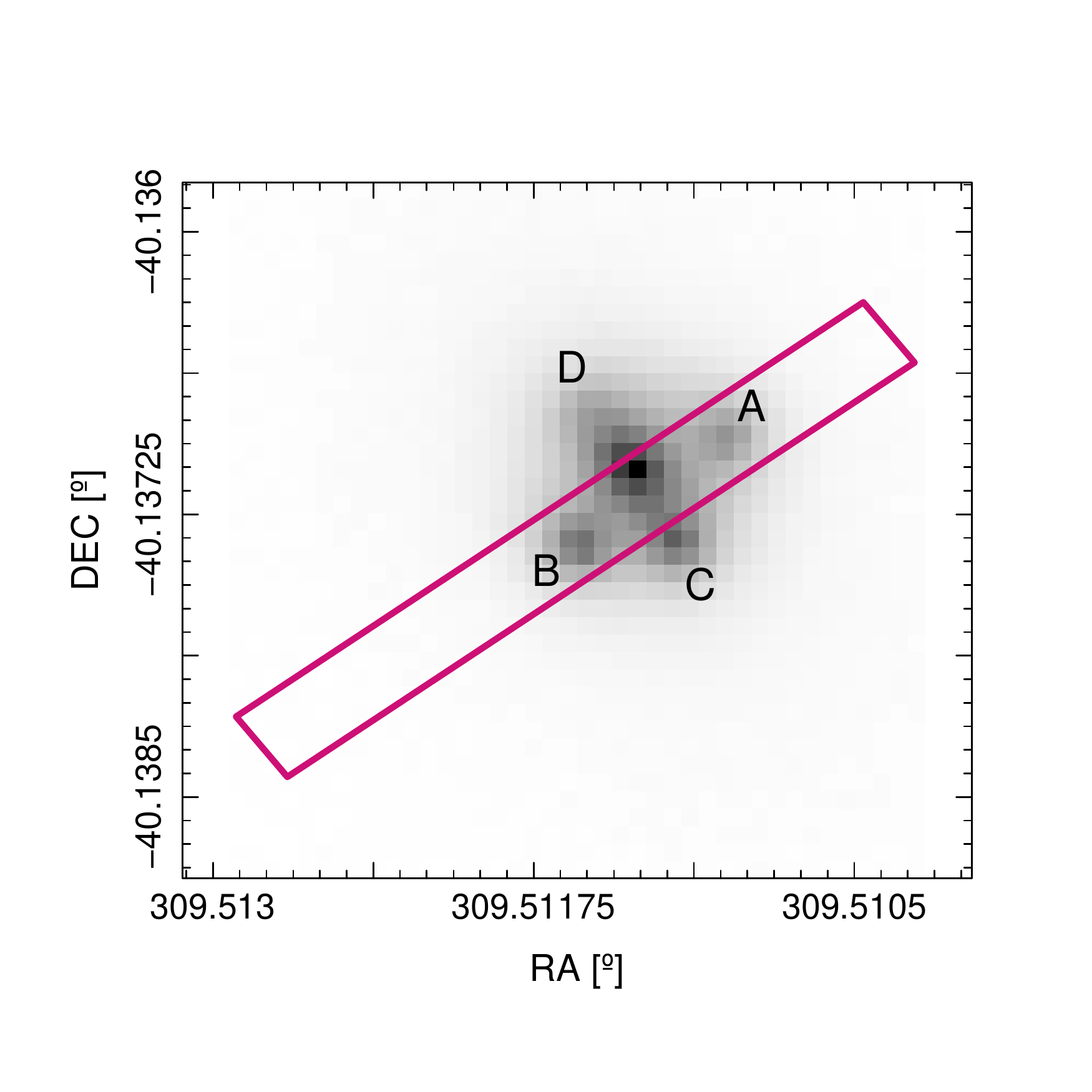}
\caption{Scheme of the slit position centered on image B (RA 307.5115875, DEC -40.13736167, epoch J2000) of $\rm WGD2038-4008$. The FITS image is from the Dark Energy Survey (DES) in filter r\protect\footnotemark.}
  \label{fig:one}
\end{figure}

We obtained spectra for $\rm WGD2038$-$4008$ during July of 2019 as part of  ESO proposal ID $\rm 103.B-0566(A)$ (PI: A. Melo) using the X-shooter instrument mounted at the 8.2 m UT2 at the Very Large Telescope (VLT), Paranal Observatory, Chile \citep{2011A&A...536A.105V}. X-shooter is a medium resolution spectrograph that observes in a wide spectral range, from ultraviolet (UVB; 3000$-$5600 \AA), through visible (VIS; 5500$-$10200 \AA) and up to the near-infrared (NIR; 10200$-$24800 \AA).  We use three observing blocks (OBs) taken in two different nights with an average seeing of 1.12\arcsec. The UVB slit was 1.0\arcsec $\times$ 11\arcsec (Spectral resolution $\rm R=5400$) while VIS and NIR slits were  1.2\arcsec $\times$ 11\arcsec ($\rm R=6500$ and 4300 for VIS and NIR) with a readout mode (UVB and VIS) of 100k/1pt/hg and a nodding of 3\arcsec per individual frame. Each UVB/VIS (NIR) OB consist of 2 (4) exposure frames. The atmospheric dispersion corrector (ADC) is used to avoid chromatic differential atmospheric refraction. Table~\ref{table:observation} summarizes the main observational characteristics of $\rm WGD2038-4008$.\\
The slit position was chosen to cover the two brightest source images, centered in image B of the gravitational lensed quasar with a position angle on the sky of $126.8514^{\circ}$ to include image A (see Figure\,\ref{fig:one}). 

\footnotetext{Gravitational Lensed Quasar Database \url{https://research.ast.cam.ac.uk/lensedquasars/}}

\subsection{Data processing \label{sec:Data_procesing}}

\begin{figure}
 \centering
  \includegraphics[width=9.0cm]{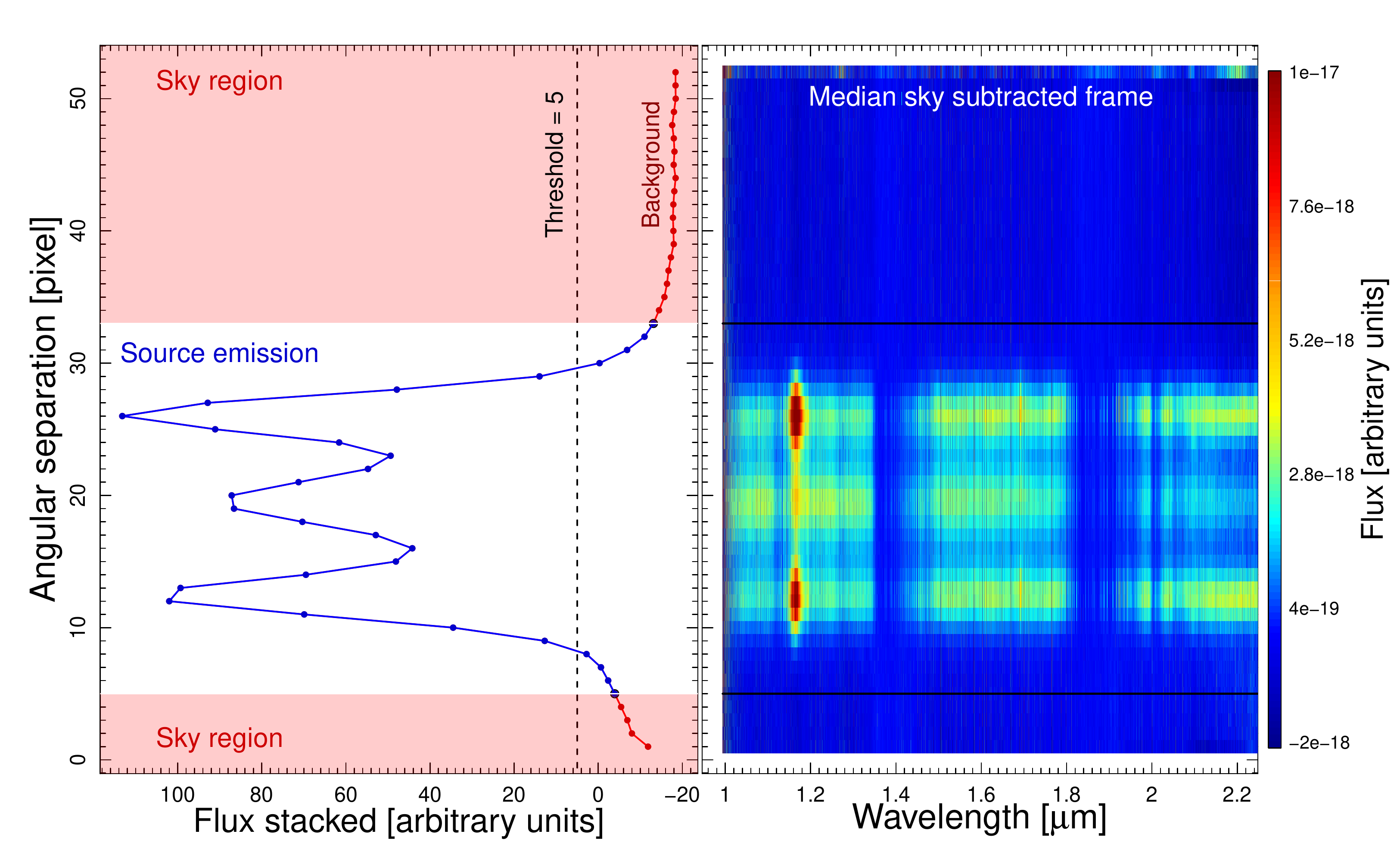}
  \caption{\textit{Left}: Collapsed frame along the wavelength axis after the subtraction of the sky median. The pink shaded zone represent the sky region (see text below). \textit{Right}: NIR frame after  sky subtraction.}
  \label{fig:exam}
\end{figure}

We used the ESO pipeline \texttt{EsoReflex} (\citealt{2013esoreflex}) workflow with the X-shooter pipeline version $3.5.0$ to reduce each OB without using nodding to subtract the sky emission. This method was employed instead of the standard one because the  3\arcsec nodding is comparable to the image separation ($\sim2.87$\arcsec), causing a self-subtraction flux from the lensed quasar spectra. The next steps  in the reduction and extraction were slightly different for the three arms of the instrument.
Once the frames were corrected by cosmetics (flat field, dark current, wavelength calibration, among others), we proceeded to subtract the sky emission in the NIR arm.
We designed a sky emission correction for each individual frame based on Principal Component Analysis (PCA, \citealt{1964DeemingPCA,1981BujarrabalPCA,1999FrancisPCA}), a method normally used in multi-dimensional analysis. PCA uses a basis of eigenvectors that are constructed to describe the data (by maximizing the variance of the projected data, for instance).  This method is usually applied to reduce the number of parameters describing a data set by computing the principal components to change the representation of the data.
The number of components used in the reconstruction was chosen to minimize the standard deviation of the residuals between the spectrum and all sky models (using different number of components).
The procedure to obtain the best representation of the underlying sky emission in each frame had the following steps. First, the outliers (such as bad pixels) were masked using $\sigma$-clipping ($\sigma$=5 with three iterations), replacing them with an estimated value obtained from a bicubic interpolation using the surrounding pixels. Then, we calculated the median for each wavelength bin to obtain a rough approximation of the sky emission as a function of the wavelength. Note that this value will only be  used to identify the targeted spectra (quasar lensed images A and B, as well as the lens galaxy). We subtract this rough sky median from each frame and collapse the remaining 2D spectra along the wavelength (see Figure\,\ref{fig:exam} right and left, respectively) to select an uncontaminated spatial region for the sky emission. A threshold equal to 3 pixels (above and below) the dispersion of the median above the background (see Figure\,\ref{fig:exam}, left) is applied to choose the region that is going to be employed as the PCA-basis (normalized to the unit). 
The PCA eigenvector basis is obtained by constructing a model of the sky emission in the selected spatial region, this 2D sky model is then subtracted from the frame.

Flux calibration is done by using the response curve  from the end-products of the X-shooter pipeline. This response is  obtained from a standard star  observed the same night as the  target, in our case GD153, EG 274 and Feige 110 for OB 1, 2 and 3,  respectively.

\begin{figure}
 \centering
  \includegraphics[width=9.0cm]{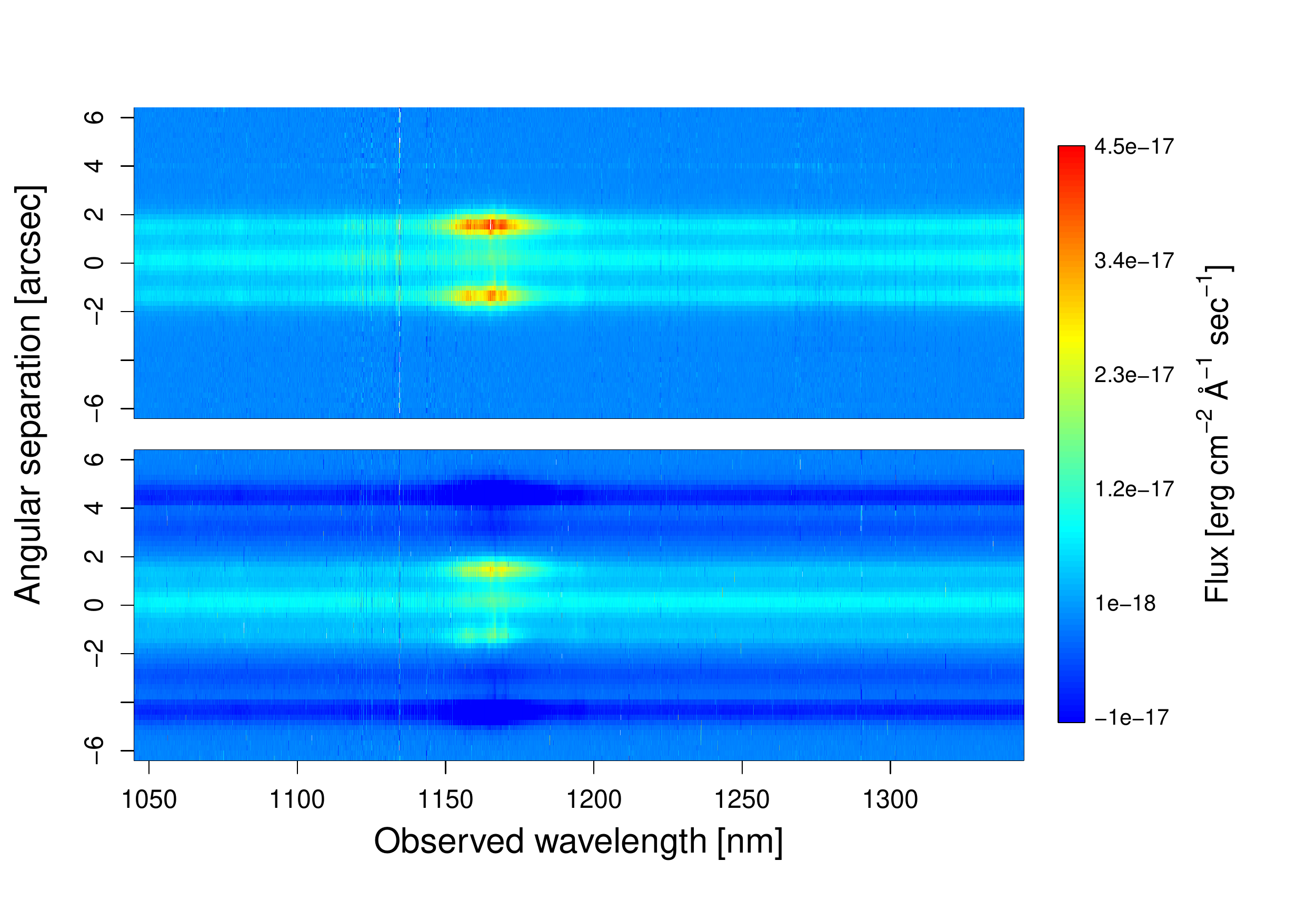}
  \caption{Final 2D NIR spectrum using PCA sky extraction described in Section\,\ref{sec:Data_procesing} (Top Figure) compared to the one using ESO pipeline reduction  (bottom figure). Both spectra are flux calibrated and telluric corrected. }
  \label{fig:two}
\end{figure}

After the sky modelling and subtraction, we  employ \texttt{molecfit} (\citealt{2015A&A...576A..77S,2015A&A...576A..78K}) in each frame to correct by telluric absorptions. For each frame, the target spectra were median-combined into a single spectrum in order to increase the signal and decrease the noise. The spatial region occupied by the targets was previously calculated  during the PCA sky emission estimation, and corresponds to the source emission region in Figure\,\ref{fig:exam}. The \texttt{molecfit} best-fit was applied to each frame row-by-row. 
Once the frames are corrected by sky emission and telluric absorption, they are median combined and the uncertainties are estimated as the median absolute deviation. All the parameters required for the stacking (dittering, pixel scale, among others) were obtained from the header of each frame, modified by the X-shooter workflow. Figure\,\ref{fig:two} shows the result of the final 2D spectrum (top panel) compared to the the one obtained as end-product from the ESO pipeline (bottom panel).

In contrast to the NIR, the VIS observations are not dominated by the sky brightness. Therefore, instead of the PCA analysis, we use the median of each sky region as the best representation of the sky brightness, obtained in the same way as for the NIR (selecting a region free of source emission to compute the sky brightness representation). The flux calibration, telluric correction, and combination of frames are done in the same way as for the NIR arm. Even though the UVB arm does not require sky subtraction, we use the same procedure as the VIS arm to be consistent with the reduction.

\subsection{1D extraction}

\begin{figure*}
 \centering
  \includegraphics[width=\textwidth]{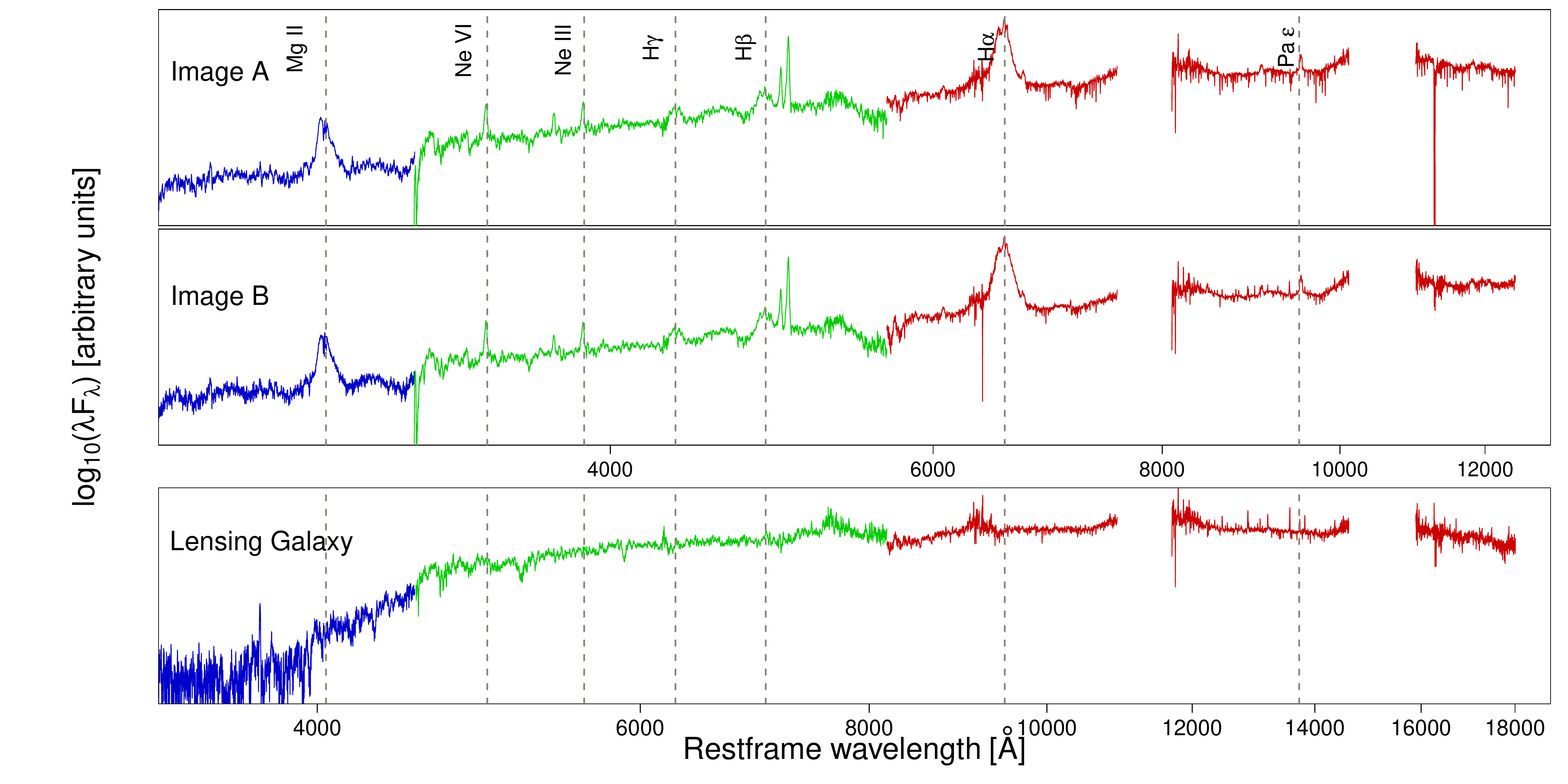}
  \caption{Rest-frame X-Shooter spectra for image A, image B and the lensing galaxy. Colors represent the different arms of the instrument: UVB (green), VIS (blue) and NIR (red). The images are corrected by their respective redshift (z$_{s}$ = 0.777 and z$_{l}$ = 0.230). The atmospheric windows are left blank for the NIR band. The lensing galaxy is uncontaminated by the quasar emission. The wavelength is in the rest frame of the respective object and the position of the emission lines are shown as dashed lines.}
  \label{fig:spectra}
\end{figure*}

Due to the small separation between the quasar images A, B and the lens galaxy, there can be cross-contamination in their spectra (see Fig.\,\ref{fig:one}). To obtain uncontaminated spectra we proceed as follows. First, we collapsed the 2D reduced spectrum along the wavelength axis in a high S/N region, for example around an emission line region (H$\alpha$ in the NIR, OIII in VIS and Mg{\footnotesize{II}} in the UVB arm). In the case of the VIS and UVB arms, we selected bins of 20 pixels (4 \AA) to increase the signal-to-noise of the sources. We masked the outliers (persistent bad pixels, poor sky subtraction and/or low signal to noise regions) to obtain the best fit parameters as a function of the wavelength. The spatial contribution of each component is estimated by simultaneously fitting  three Gaussian profiles. The distances between image $A$ and $B$ ($\sim$2.87 arcsec $\approx$ 14 pixels), and between image A and the lens galaxy projection onto the $AB$ segment ($\sim$ 1.47 arcsec $\approx$ 8 pixels) are used to fix the position of the Gaussian centers for $B$ and the lens galaxy, respectively. Assuming $A$ and $B$ are point sources, we can consider that they have the same FWHM (and standard deviation parameter - $\sigma$), and a variable $\sigma_l$ (larger than $\sigma$, due its extended emission) for the lensing galaxy in UVB and VIS arms. Due to the faintness of the lens galaxy in the NIR arm,  the $\sigma_{l}$ value of the lens galaxy is considered the same as for the images ($\sigma_{l}=\sigma$). Thus, the free parameters are the amplitudes, image $A$ center, $\sigma$ for the point sources (and lens galaxy in NIR) and $\sigma_l$ for the lens galaxy (in the UVB, VIS arms).
Using the best-fit estimated values and their respective uncertainties, we construct a probability function for the spatial distribution of each target (quasar images and lens galaxy), allowing to identify the probability that a given spatial pixel belongs to one of the targets. We use error propagation for each free parameter to estimate the related uncertainties in each final, uncontaminated spectra.\\
Considering the seeing variation along the wavelength range and the selected slit width, we need to estimate the percentage of lost flux. We estimated the broadening of the spectra profile due to the instrumental dispersion at different wavelengths by fitting a Gaussian function for each wavelength bin of the telluric standard star (HD 115470 for the case of OB 1 of seeing 0,75\arcsec). The $\sigma_\odot$ obtained was of 0.76\arcsec with a dispersion that does not vary from 1 pixel between the arms.
We use this value (see in Table\,\ref{table:observation}) to calculate the percentage of flux entering the slit by simulating the system as a sum of Gaussian functions and sigma obtained from the header and integrating it within the slit using the seeing delivered in the header. The percentage of flux lost was 30.5$\%$ in UVB, 14.9$\%$ in VIS and 19.74$\%$ in NIR. These values will be included as an extra flux error in all our analysis.\\
After extracting the spectra for the three components, we notice that the lens galaxy spectrum shows contamination by quasar emission lines. Given the amount of contamination, the slit width and position angle, we infer this is a contribution from image C (see Fig.\,\ref{fig:one}). To obtain the uncontaminated lens galaxy spectrum, we use spectrum A as a proxy for C  and estimate the C contribution fraction for each arm ($0.25$ for NIR, $0.43$ for VIS and $0.5$ for UVB arm) that removes the quasar emission lines.
The S/N of the emission line and continuum were obtained by estimating the standard deviation of the background.
We use the 2D spectra to obtain the background emission (sky emission in Section\,\ref{sec:Data_procesing}), getting the mean and standard deviation of this background.
We then choose a continuum region located around the emission line (50 \AA) for each signal contribution (image A, B and the lensing galaxy) and obtain the mean value. With these values we calculate the S/N of the continuum. We calculate the S/N of the emission lines using the same method, i.e. selecting the same spatial region but estimating the mean in a reduced wavelength window (approximately 300 to 500 \AA) around each emission line. The S/N for each continuum and emission line are listed in Table~\ref{table:tablembh}.\\
The final spectra for images A, B and the lensing galaxy (uncontaminated by the quasar emission) is presented in Fig.\,\ref{fig:spectra}. As the ADC did not work during the night that OB 1 and 2 were taken,  the UVB and VIS arm experienced flux loss (Fig.\,\ref{fig:spectra}), explaining the atypical profile of the AGN spectra (see  \citealt{2001vandenberk} and \citealt{2006Glikman} for a composite quasar spectra). This loss will affect the luminosity measurement, specially in the UVB arm (See section \ref{sec:emissionandlum} for more details).

\begin{figure*}
  \includegraphics[width=0.46\textwidth]{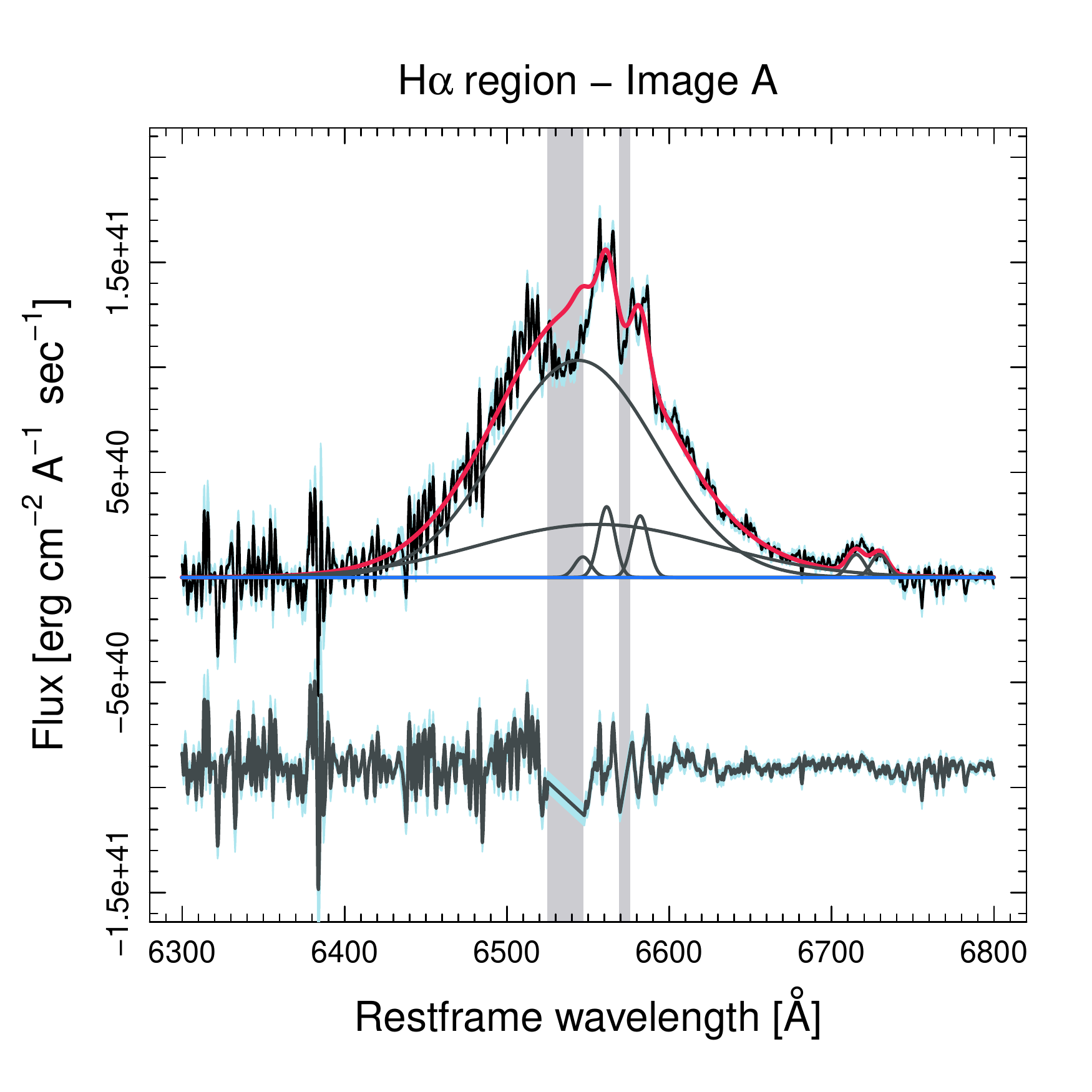}
  \includegraphics[width=0.46\textwidth]{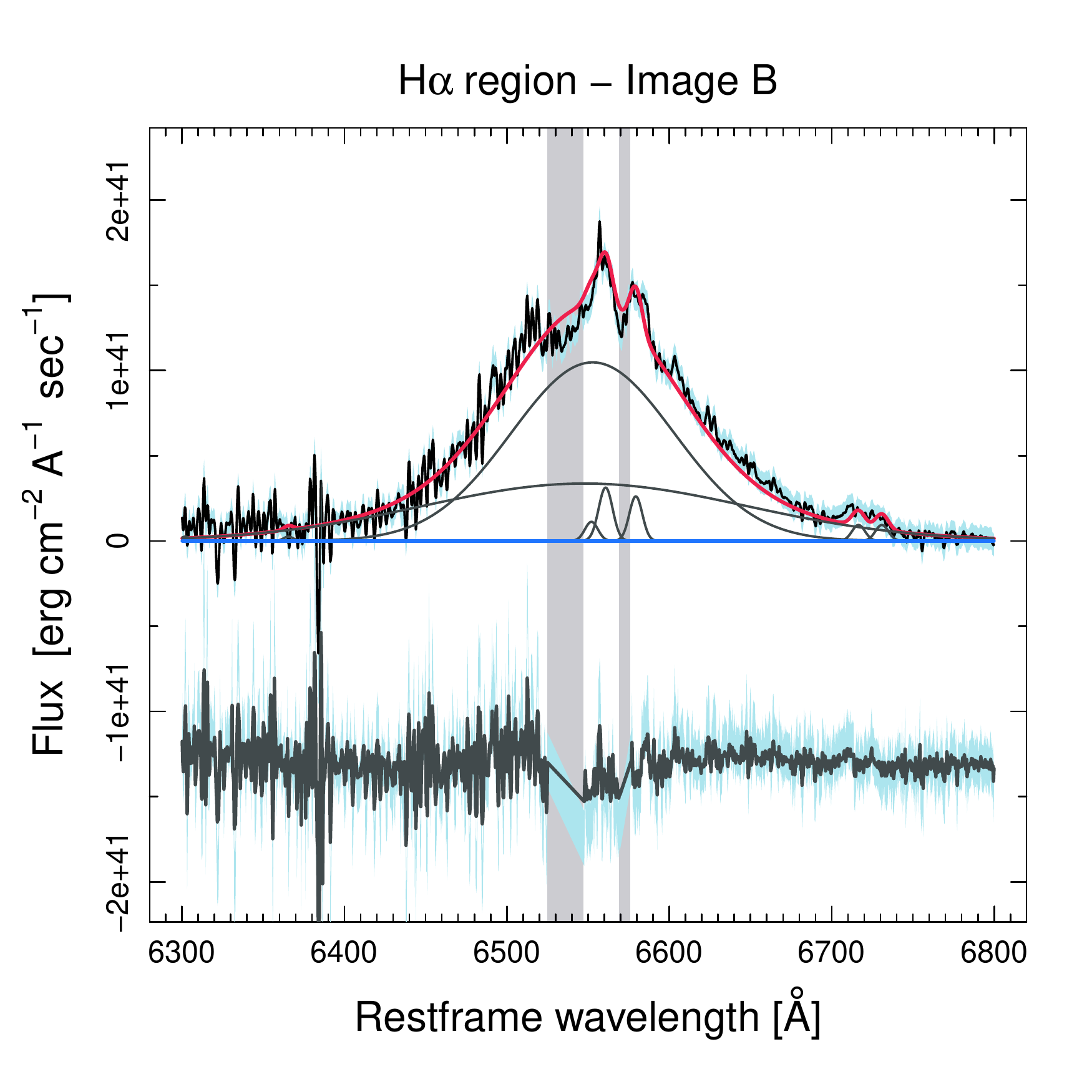}\\
  \includegraphics[width=0.46\textwidth]{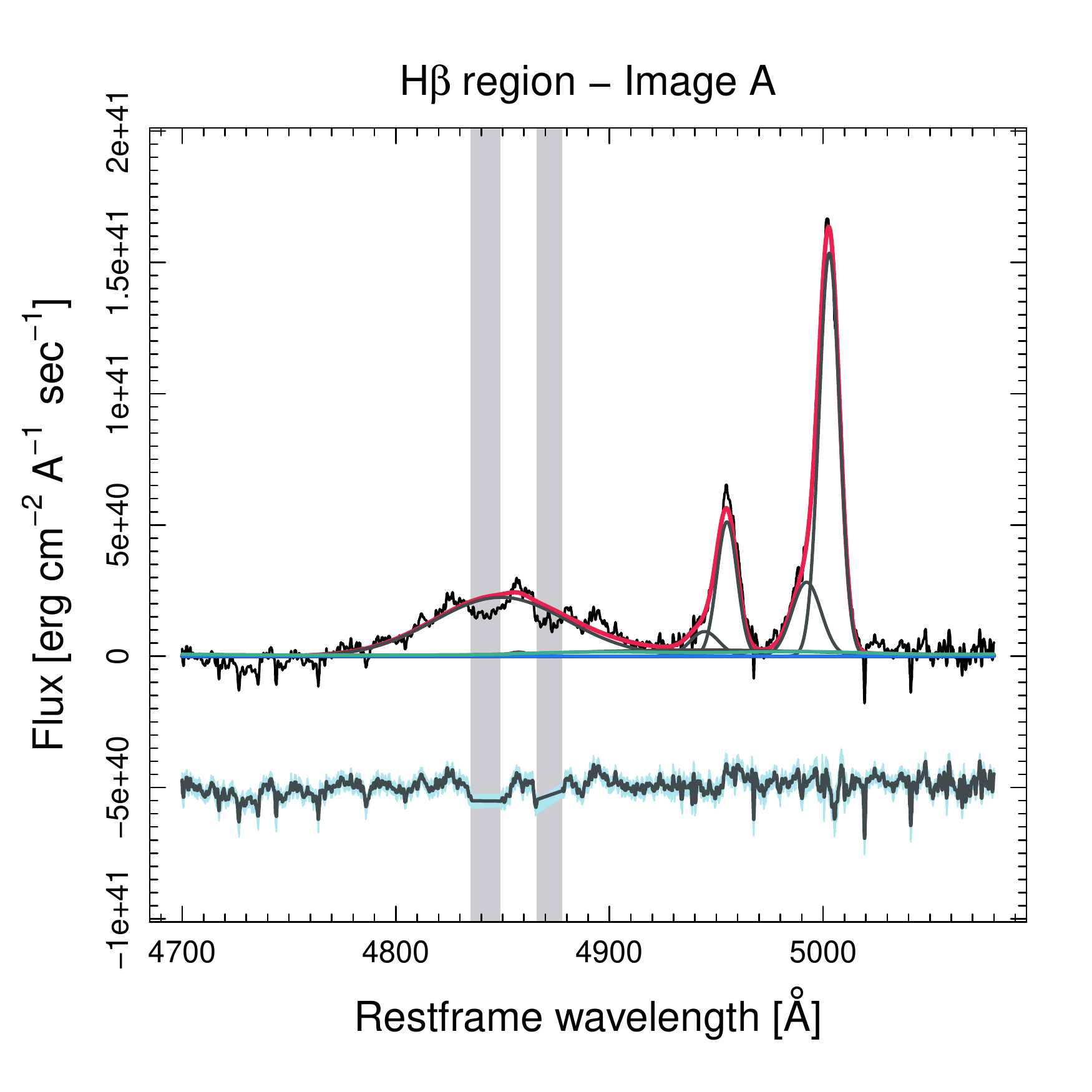}
  \includegraphics[width=0.46\textwidth]{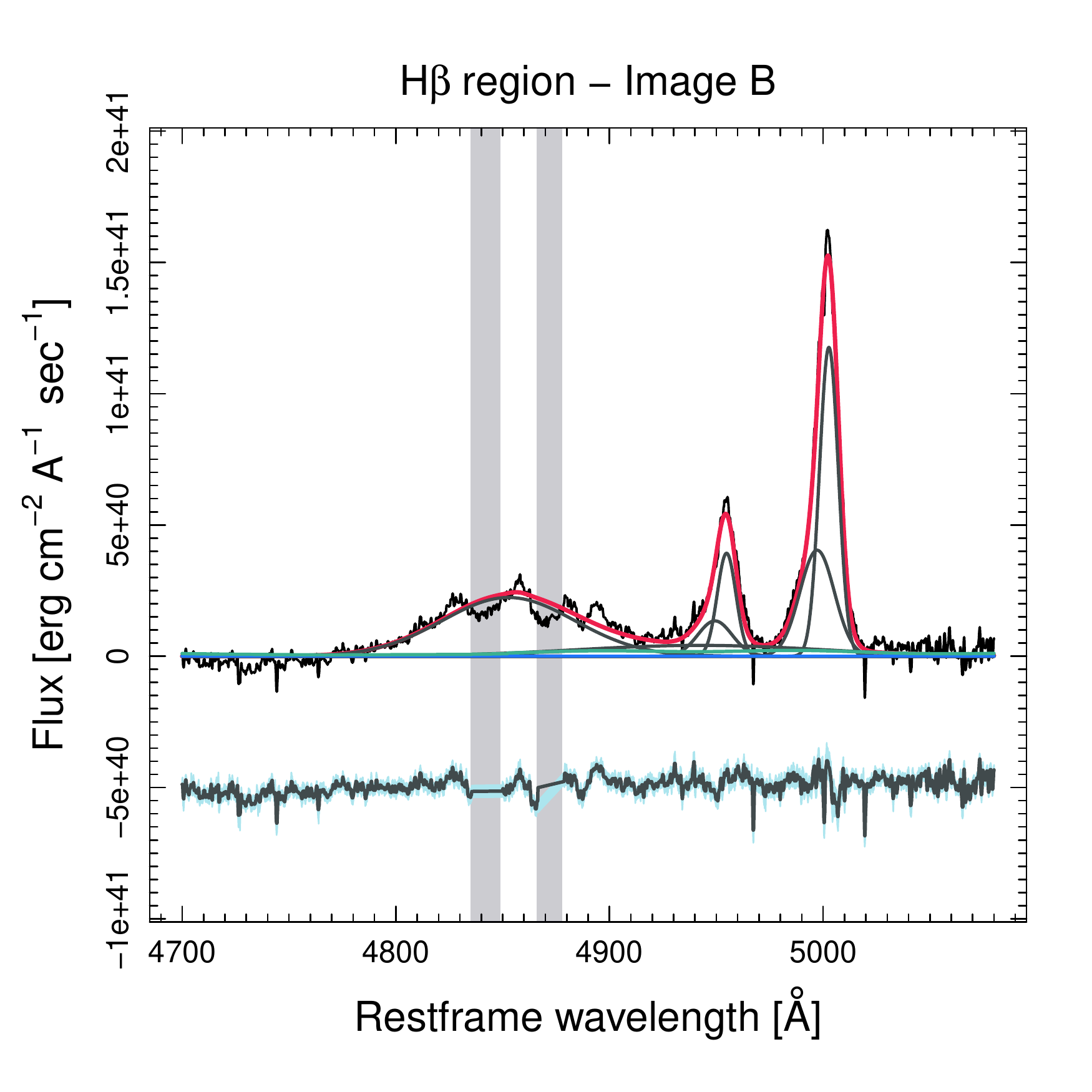}\\
  \includegraphics[width=0.46\textwidth]{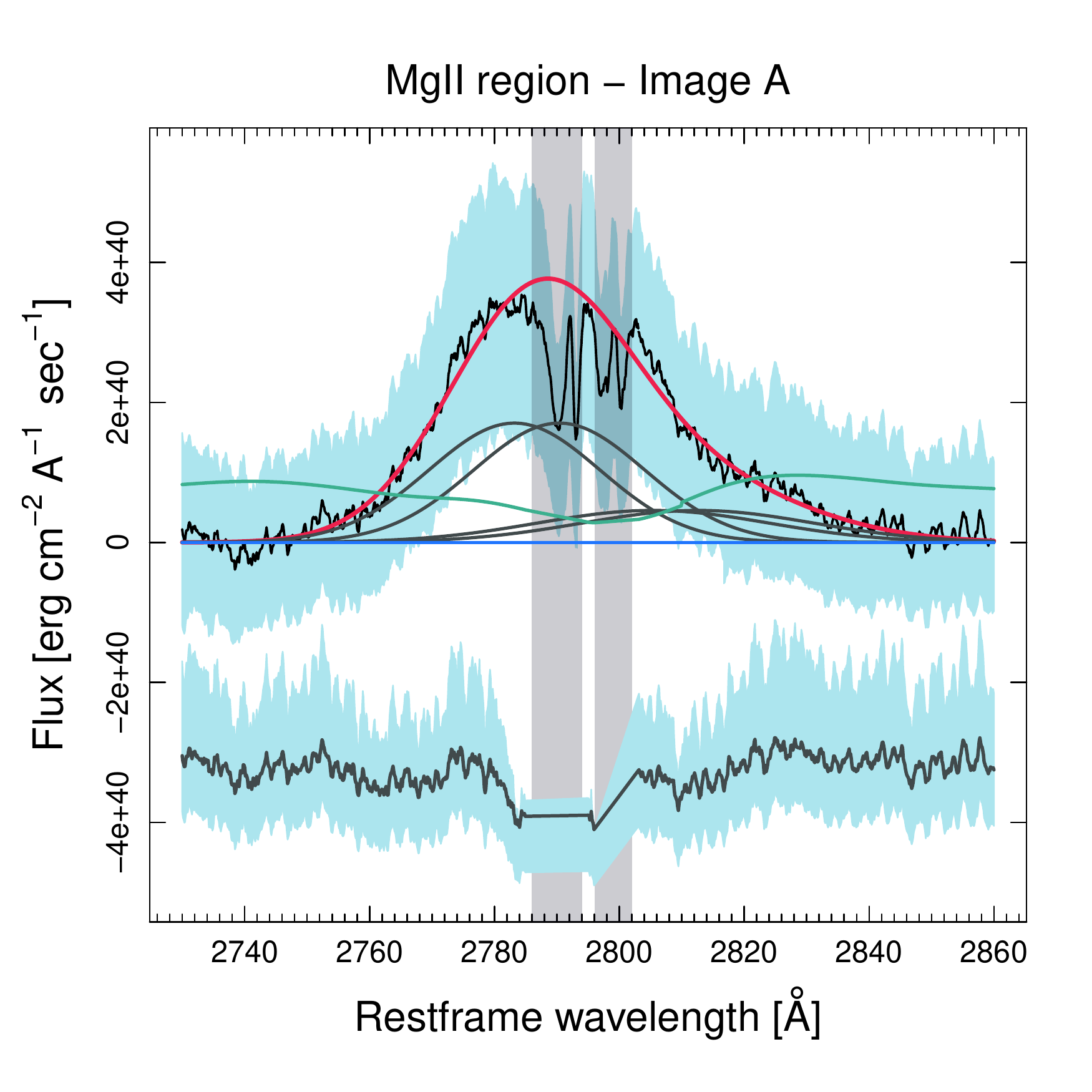}
  \includegraphics[width=0.46\textwidth]{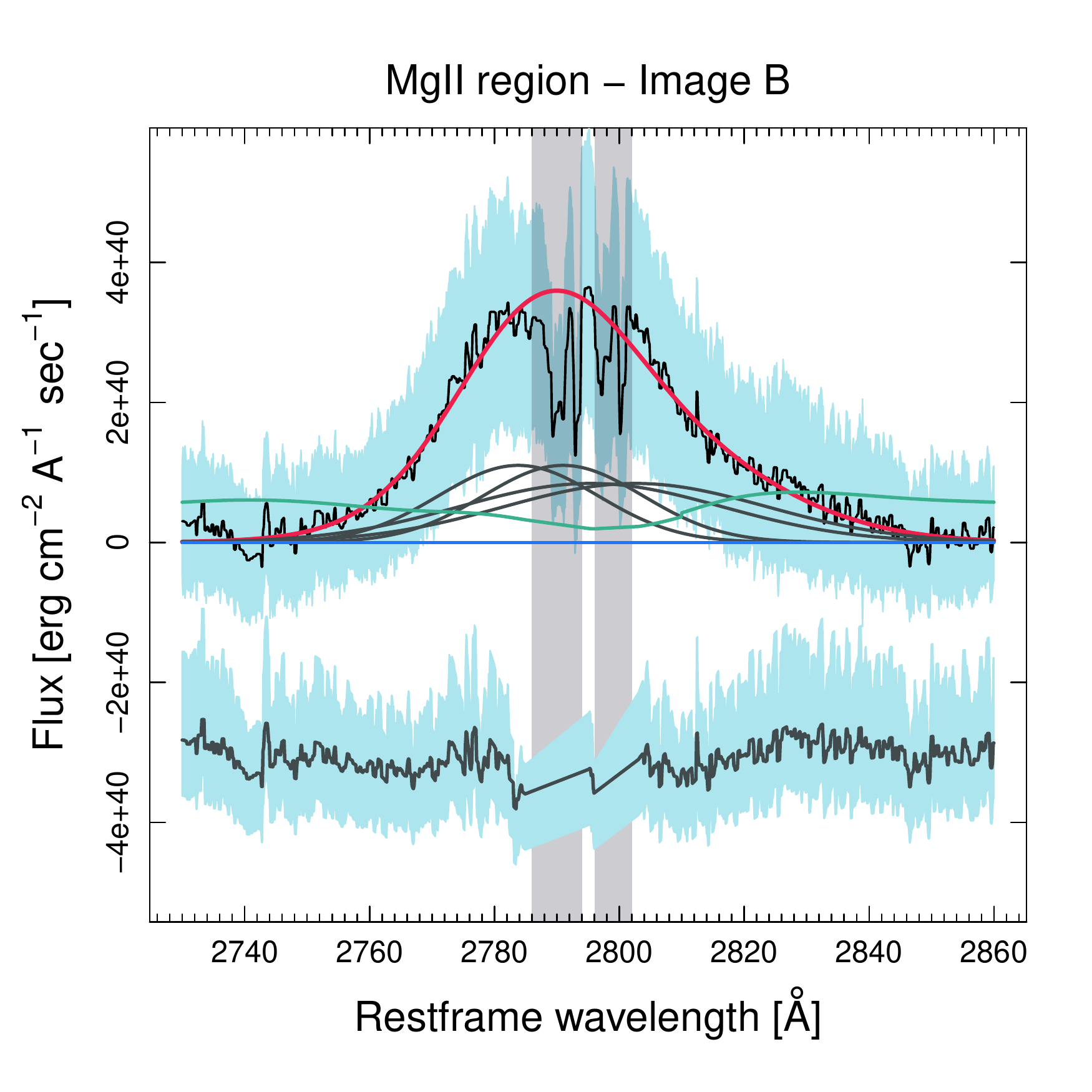}\\
  \caption{Gaussian fitting of H$\alpha$, H$\beta$ and MgII regions for images A (left) and B (right). Red line represents the best fit, black lines represent the different components of each region (emission and absorption), green line represents the Fe template and the blue line is the continuum fit of the spectra. 1-sigma error of the spectra along with the residuals and their respective errors are in the bottom of the images.
  }
  \label{fig:mbh}
\end{figure*}

\begin{figure}
 \centering
  \includegraphics[width=8.0cm]{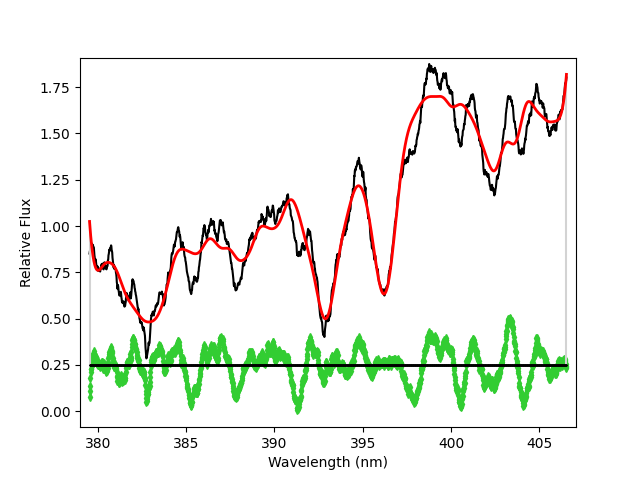}
  \caption{Fit of stellar templates to the lensing galaxy using pPXF package after removing the extra quasar contribution.}
  \label{fig:disp}
\end{figure}

\begin{figure*}
 \centering
  \includegraphics[width=\textwidth]{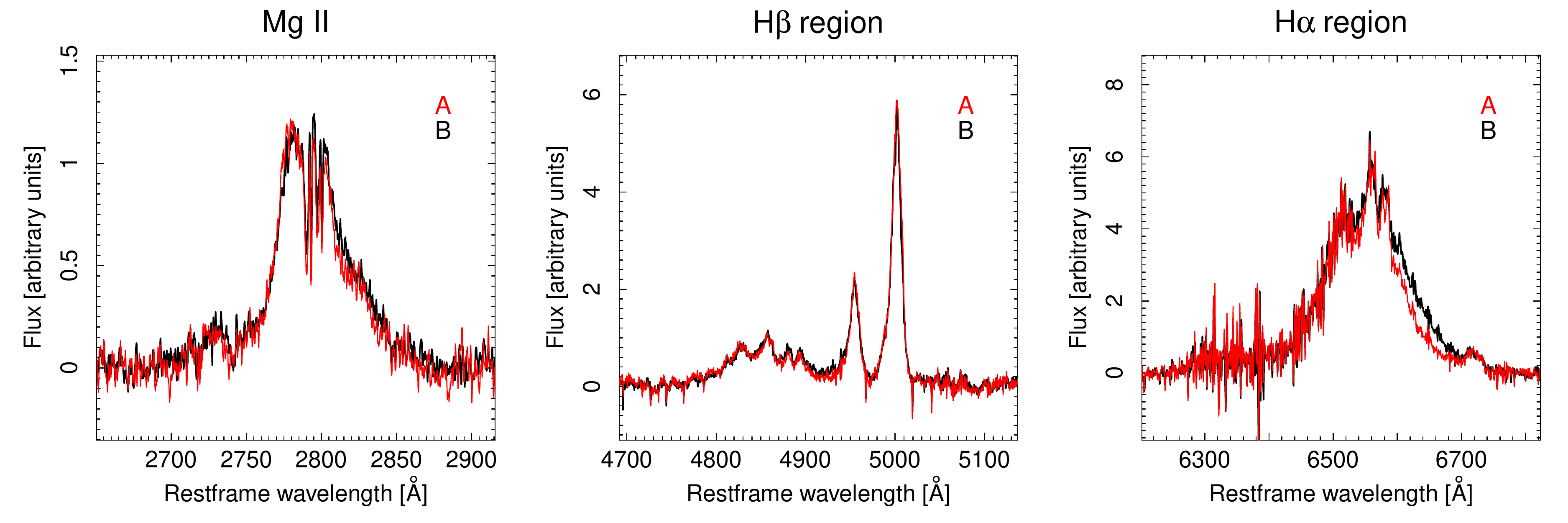}
  \caption{Mg{\footnotesize{II}}, H$\beta$ and H$\alpha$ emission line region. The images are demagnified using the magnification values given in section\,\ref{sec:Method}. }
  \label{fig:perfil}
\end{figure*}

\begin{table*}
\caption{Log of the observation for the three OBs.}             
\label{table:observation}      
\centering
\begin{tabular}{ l c c c c l c } 
\hline\hline 
OB & Date & Arm & Exp. time (seconds) & Number of Exposures & Airmass & Seeing\tablefootmark{1} ($\arcsec$) \\
\hline
1 & 9/10 July 2019	  & UVB & 600 & 2  & 1.067 & 1.02  \\
                    & & VIS & 600 & 2  & 1.068 & 1.09  \\
                    & & NIR & 600 & 4  & 1.065 & 1.08  \\
\hline
2 & 9/10 July 2019	  & UVB & 600 & 2 & 1.133 & 1.08  \\
                    &     & VIS & 600 & 2 & 1.134 & 1.02  \\
                    & 	  & NIR & 600 & 4 & 1.127 & 1.01  \\
\hline
3 & 10/11 July 2019  & UVB & 600 & 2 & 1.133 & 1.27  \\
                    &     & VIS & 600 & 2 & 1.134 & 1.30  \\
                    & 	  & NIR & 600 & 4 & 1.127 & 1.28  \\
\hline
\end{tabular}
\tablefoot{
\tablefoottext{1}{This seeing corresponds to the delivered seeing on Image Analysis detector given in the header of each frame.}}
\end{table*}

\section{Methods\label{sec:Method}}

Black hole mass is estimated by using the single-epoch method (SE method; e.g. \citealt{2004MclureDunlop,2008Shen,2012trakhtenbrot}), which combines the usage of the Doppler line width of the broad emission line and the monochromatic luminosity to obtain a proxy for M$_{BH}$. If we assume that the emitting gas in the BLR is virialized:

\begin{eqnarray}
M_{\text{BH}} &=  f \,  R_{BLR} \,  (\Delta v)^2 ~G^{-1}, 
\end{eqnarray}

\noindent where $G$ is the gravitational constant, $R_{BLR}$ is the BLR size, $(\Delta v)^2$ is the velocity of the line emitting gas in the BLR, and $f$ is the virial factor that depends on the unknown structure, kinematics, inclination and distribution of the BLR (\citealt{2004Peterson} and references therein). The BLR size comes from the Reverberation Mapping (RM, \citealt{1982Blandford,1993Peterson}) and with the known correlation between the  AGN luminosity and the size of the BEL, $R_{BLR} \sim (\lambda L_{\lambda})^{\alpha}$ (e.g. \citealt{2000kaspi,2005kaspi,2009Bentz}), allowing us to estimate M$_{BH}$ as:

\begin{eqnarray}
\log( M_{\text{BH}}) &= \log(K) \,  + \, \alpha \, \log\left(\frac{\lambda L_{\lambda}}{10^{44} \text{ erg/s }} \right)\nonumber \\
 &  + \, 2.0 \, \log \left(\frac{\text{FWHM}}{1000 \:\text{km/s}} \right),    \label{eq:two}
\end{eqnarray}

\noindent where $K=G^{-1}f$. 
Although, the literature shows different values for the parameters $K$ and $\alpha$ (\citealt{2004MclureDunlop,2006Vestegaardandeterson,2009vestergaardandosmer,2011shen}), we use those estimated by \citealt{2016MNRAS.460..187M} because they were estimated using a similar observing setup, thus minimizing the systematic effects. In particular, the sample of \citealt{2016MNRAS.460..187M} not only contains several emission lines for each object but also all lines for a single object were observed simultaneously. The values for the parameters used for the emission lines (H$\alpha$, H$\beta$ and Mg{\footnotesize{II}}) at their respective luminosities (L$_{5100}$, L$_{5100}$, L$_{3000}$) are:\\

\noindent
( log $K$ , $\alpha$ ){\textbf{$_{H\alpha}$}} = ( 6.845 , 0.650 )\\
( log $K$ , $\alpha$ ){\bf $_{H\beta}$} = ( 6.740 , 0.650 )\\
( log $K$ , $\alpha$ )$_{Mg{\footnotesize{II}}}$ = ( 6.925 , 0.609 )\\

Besides the usual uncertainties  related to the SE method (FWHM, luminosity, and $f$ parameter estimations), the observed source luminosity needs to be corrected for the lensing magnification. To obtain the magnification factor ($\mu$) we use the convergence ($\kappa$) and shear ($\gamma$) parameters estimated from the lens model as $\mu = 1/ [ (1- \kappa)^2 - \gamma ^2 $] (\citealt{1996Narayan}).
Employing the values previously calculated by \citealt{shajib2019}, we obtain a magnification factor of  $\mu_A$ = 2.27 $\pm$ 0.21 and $\mu_B$ = 2.71 $\pm$ 0.32. 

\subsection{Emission line fitting and luminosity measurement \label{sec:emissionandlum}}
After demagnifying the spectra, and removing the continuum and the iron template, (following \citealt{2016MNRAS.460..187M}), we model each emission line profile and estimate the BEL FWHM. We used Gaussian functions to represent the broad and narrow components of each emission line (see Table 4 of \citealt{2007shang}) and mask those regions affected by absorptions\footnotemark. \footnotetext{We should point out that we also considered other line profile fittings: i) Gaussian fitting without masking regions, ii) the addition of Gaussian profiles for the absorption features. Although both methods provide FWHM that are consistent with our results, the first one yields larger residuals and the last one introduces overfitting.} In the H$\alpha$ region we added four extra Gaussian components for [N II] and [S II] NLR doublets. In the H$\beta$ region we considered two extra Gaussians for the [O III] NLR doublet and one for the He II broad emission line. For the Mg {\footnotesize{II}} region we considered two narrow and broad Gaussian components.
The FWHM used for the M$_{BH}$ measurement is obtained from the standard deviation of line profile after removing the NLR component (i.e. the resulting profile is the combined Gaussians representing the broad line components). The uncertainties were obtained using error propagation and a Monte-Carlo simulation of 1000 random resamplings, assuming a Gaussian distribution for the flux uncertainty at each pixel. The best line fit is shown in red in Figure~\ref{fig:mbh}. As every emission line exhibits some kind of distortion (possible absorptions) that could lead to an overestimation of their FWHM, we decided to mask those regions for the Gaussian fitting.
Mg {\footnotesize{II}} profile has several absorption features (masked region [2787:2794 , 2796:2802]~\AA), possible caused by the circumgalactic medium (CGM). In the case of H$\alpha$, the feature perceived as absorption masked region [6525:6547 , 6569:6576]~\AA~ could be instead a very bright NLR. Similarly, H$\beta$ shows a distorted profile (masked region [4835:4849 , 4866:4878]~\AA) possibly related to a poor FeII fitting. 
Monochromatic luminosity was measured using continuum windows on each side of the emission line ( [4670 : 4730 , 5080 : 5120] \AA \,  for H$\beta$ and [6150 : 6250  , 6800 : 6900] \AA \, for H$\alpha$). 
These spectral windows have been selected due the small (or even null) emission line contamination, and were used to interpolate the region of interest following a single power-law function. 
As mentioned before, the flux loss in the UVB and VIS arms impede the use of the spectra to estimate the monochromatic luminosity at 3000 \AA. Instead, we estimated this luminosity by fitting a Spectral Energy Distribution (SED; template of \citealt{2010Assef}) to the unmagnified magnitudes obtained from HST (\citealt{shajib2019}) and DES (\citealt{Agnello2018}). 
Compared to the luminosity measurement, the FWHM of Mg{\footnotesize{II}} FWHM is not affected by this flux loss.\\
The FWHM obtained from the line profile fitting and the monochromatic luminosity (estimated from the continuum and SED) for each emission line in image A and B is presented in Table~\ref{table:tablembh}.

\subsection{Microlensing analysis}

Microlensing can induce flux variations in the quasar images due to lensing from stars in the lensed galaxy halo (see for instance \citealt{1979chang,2006wambsganss}). This flux variation in one (or more) images is sensitive to the angular size of the source, meaning that the magnification will be bigger for a smaller emitting region.
Given the latter, we could study the inner structure of WGD2038-4008 from the single-epoch images of different observations, where the accretion disk and BLR can be affected differently by microlensing and could affect the wings of the emission lines. On the contrary, the Narrow Line Region (NLR) is insensitive to microlensing and can be used as the baseline (\citealt{2002Abajas}).
To investigate whether microlensing is present, we use the magnitude difference between the emission line core and the continuum (see \citealt{2003Moustakas,2009ApJ...706.1451M,2011mediavilla,Motta2012,Guerras2013,Motta2017,Rojas2020}). 
The quasar emission lines have different components, meaning that they come from different inner regions of the AGN, the line core will be dominated by the NLR while the wings are dominanted by the BLR emission. Also, the light of each image follows a different path through the lens galaxy where gas and dust can produce extinction. 
We can separate microlensing from extinction considering that only the latter will affect both the continuum flux ratio and the core of the emission line (\citealt{1999Falco,2002motta,2005Mediavilla}). 
We obtain the magnitude difference between components in the continuum by fitting a straight line between two regions in each side of each emission line. The regions have a size of $\sim 30-40  \AA$. We integrated the line between the two regions to obtain the continuum flux for both images. This continuum is then subtracted from the spectrum and we integrate a small window (between 10-30 $\AA$) centered in the emission line core to obtain the flux uncontaminated by the continuum. Integrating in the core of the emission line decreases the BEL contamination that can also be affected by microlensing.
The uncertainty of the flux is assumed to be related to the continuum fitting and is obtained using error propagation, where the square root of the error in the spectra and the straight line are added in quadrature. Finally, the magnitude difference between image A and B for each region (continua and line cores) is $\rm  m_A - m_B =-2.5\log (F_A/F_B)$, obtaining a $\rm (m_A - m_B)^{line}$ for the core of the emission line and a $\rm (m_A - m_B)^{cont}$ for the continuum.

\subsection{Stellar velocity dispersion of the lens galaxy}

The stellar velocity dispersion ($\sigma$) of a galaxy measures the random motion of stars due to a presence of a mass. Obtaining an accurate value of $\sigma$ is important to restrict the lens model parameters, and together with the light curves of the images of the lensed quasar measure the Hubble Constant $H_0$ and helps to improve the uncertainties. The velocity dispersion was estimated from the lens galaxy spectrum using the Penalized Pixel-Fitting (pPXF;  \citealt{2004cappellarippxf,2017cappellarippxf}). We use the rest-frame wavelength 3600 to 4200 \AA~ for the UVB arm and 4800 to 5800 \AA~ for VIS arm. The spectra was fitted using the Single Stellar Population library by \citealt{2010vazdekis} included in pPXF (see Figure~\ref{fig:disp}). The velocity dispersion obtained was $299\pm12~ \text{km/s}$, consistent with the measurements of \citealt{Buckley-Geer2020} ($296\pm19~  \text{km/s}$ and $303\pm24~ \text{km/s}$ using Gemini South/GMOS-S spectra).

\begin{figure}
 \centering
  \includegraphics[width=8.5cm]{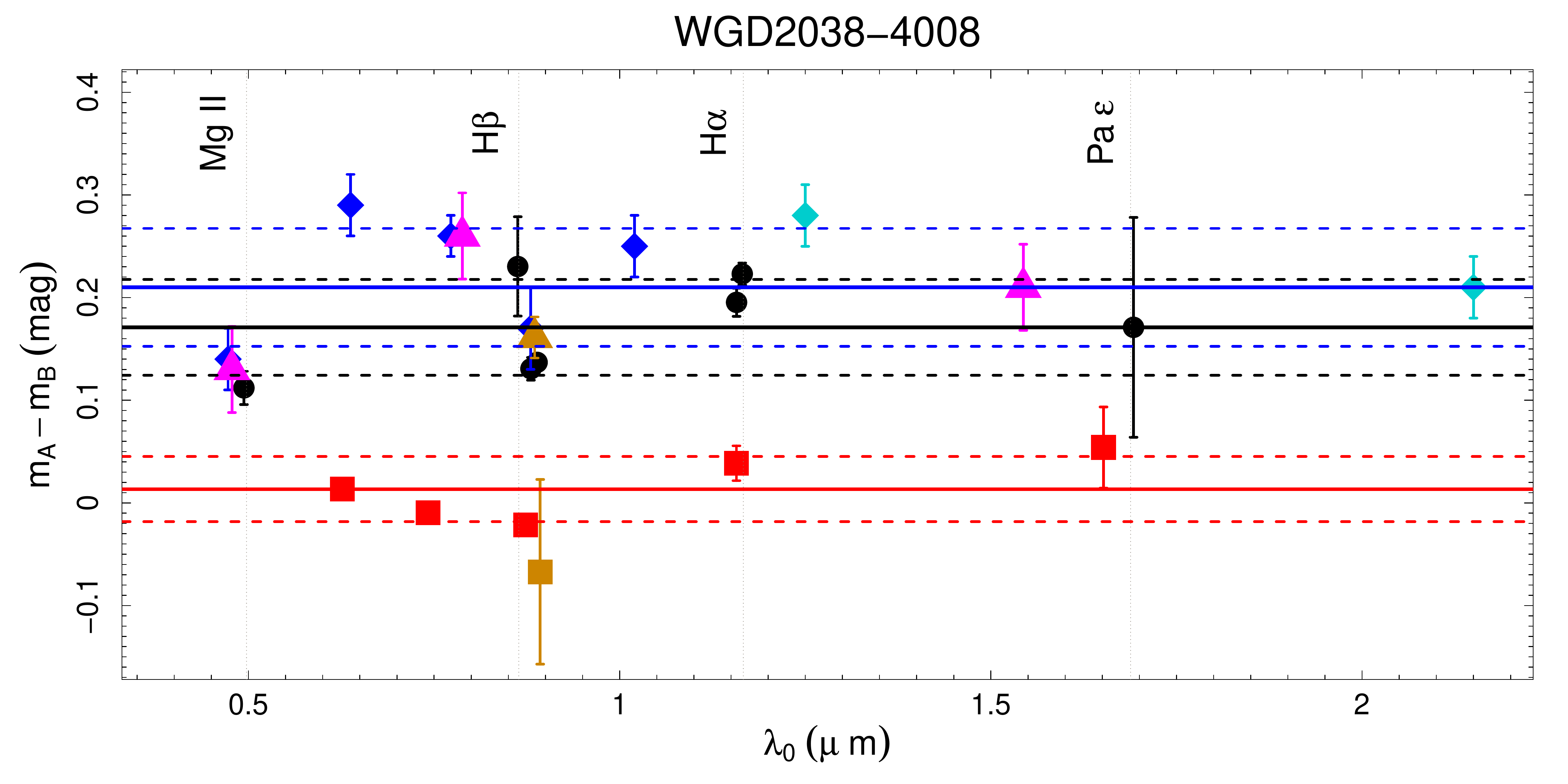}
  \caption{Magnitude difference m$_A$ - m$_B$ vs $\lambda_0$ between images A and B. Red squares shows the integrated continuum and in black circles the emission line core without the continuum using X-shooter. We include measurements obtained from the literature: HST \citealt{shajib2019} (magenta triangles), VISTA \citealt{Lee2019} (cyan diamond), DES \citealt{Agnello2018} (blue diamond) and HST  F105W/G102 \citealt{Nierenberg2020} (orange square for continuum and orange triangle for a narrow emission line). The Red line is the median of the continuum and the dotted red line the standard deviation, the black line for the emission line core and the blue line for the literature. }
  \label{fig:micro}
\end{figure}

\section{Results\label{sec:results}}
 
We identified the three most prominent emission lines of the lensed quasar (Mg{\footnotesize{II}}, H$\beta$ and H$\alpha$) with high S/N (see Table~\ref{table:tablembh}). The spectra were de-magnified using the parameters from the lens model of \citealt{shajib2019} and the continuum was subtracted to compare the profiles of image A and B (Fig.\,\ref{fig:perfil}). Interestingly, we find an enhancement of the right wing of H$\alpha$ emission line of image B compared to image A (between $\sim$ 6600 and 6700 $\AA$). This magnitude difference ($\sim 0.28 \pm 0.03$ mag integrated in the region [6591.4:6686.5] \AA) could be explained assuming that microlensing is affecting the H$\alpha$ broad emission line. This effect should also be seen in the H$\beta$ profile as it arise from a similar region size as H$\alpha$. However, we do not detect this effect, although this could be due to the low S/N of H$\beta$ ($\rm S/N \lessapprox 20$) compared to H$\alpha$ ($S/N \gtrapprox 72$) and to the presence of absorption-like features. There is no sign of such effect in the Mg{\footnotesize{II}} profile (S/N $ \lessapprox 31$), which is reasonable because Mg{\footnotesize{II}} emission is produced in a farther away region than the Balmer lines (\citealt{2012Goad}) and hence is less susceptible to microlensing effects. \\
Multi Gaussian fitting of image A and B are shown in Figure~\ref{fig:mbh} following the procedure described in Section \ref{sec:emissionandlum}. The grey shaded regions represent the masked sections used during the fitting. Table~\ref{table:tablembh} shows the FWHM estimated for each broad emission line in each quasar image.
Even though the components of Mg{\footnotesize{II}} have a slight different amplitudes due to the absorption that are contributing to the profile, the FWHM values are within the errors. The FWHM of H$\beta$ are in good agreement in spite of the low S/N. In the case of H$\alpha$,the estimated FWHM are different ($>$5 sigma) and we will discuss below how this might affect our M$_{BH}$ estimations. 
To investigate whether microlensing is present in the continuum, we obtained the magnitude difference between the core of each emission line uncontaminated by the continuum, $ \rm (m_A - m_B)^{line}$, and the continuum under the emission line, $\rm (m_A - m_B)^{cont}$, shown in Fig.\,\ref{fig:micro}. The H$\alpha$ emission line region shows two values corresponding to two emission line peaks, avoiding the right wing (see Figure\,\ref{fig:perfil}), integrated in the windows [6500.1:6526.0, 6547.7:6565.4] \AA, respectively. The H$\beta$ region shows three values corresponding to H$\beta$ ( [4820:4890] \AA~ integration window) and the [OIII] doublet emission line cores ( [4949.2:4960.0, 4996.4:5007.0] \AA~ integration window). We included the Mg{\footnotesize{II}} region integrated between [2774.6:2784.1] \AA~ and Paschen$\epsilon$ [9512.0:9533.0] \AA. Considering that the magnitude difference in the emission lines is approximately constant, we use the median and its standard error, $\rm (m_A - m_B)^{line} = 0.17 \pm 0.05$ mag, as our baseline of no-microlensing. As the values for the integrated continuum also yield a roughly constant value along the wavelength, we use the median to estimate $\rm (m_A - m_B)^{cont} = 0.01 \pm 0.03$ mag. We compare our result with spectroscopic data of the integrated flux obtained by \citealt{Nierenberg2020} (see Fig.\,\ref{fig:micro}) $\rm (m_A - m_B)^{cont}_{N} = -0.06 \pm 0.09$ and $\rm (m_A - m_B)^{line}_{N} = 0.16 \pm 0.02$,  which is in agreement with our magnitude difference for the continuum and emission line in the H$\beta$ region, respectively. Published data from broadband photometry (\citealt{Agnello2018,Lee2019,shajib2019} taken between 2016 and 2017) is also included in the Fig.\,\ref{fig:micro}. We fit a median function to these values obtaining $\rm (m_A - m_B)^{lit} = 0.21 \pm 0.06$ mag. The values are in agreement with the core of our narrow emission lines.\\ 
Since the above mentioned data are not time-delay corrected, the magnitude difference estimated from our spectra,
$\rm \Delta m = (m_A - m_B)^{cont} - (m_A - m_B)^{line} = -0.16 \pm 0.06 ~$ mag., could be due to intrinsic variability coupled with a time lag between the images. We use the \citealt{2008Yonehara} procedure to estimate such effect. We assume the structure function inferred from imaging data of quasars (\citealt{2004vandenberk}), an absolute magnitude range for the source in I band, M$_{\rm I}$= (-21,-30) , the predicted time delay for the quasar images $\Delta t_{\text{AB}} = -6 \pm 1$ days (\citealt{shajib2019}), and assume no lag between our observations as they were obtained with 1 day of difference. We obtained a magnitude difference induced by time delay coupled with intrinsic variability of 0.05~mag (0.03~mag) to 0.07~mag (0.04~mag) for a -21.~mag (-30.~mag) source in the F160W and F475W HST broad-band filters, respectively. On the other hand, we also use light-curves obtained by COSMOGRAIL (e.g. \citealt{2017bonvin,2011Courbin,2005Eigenbrod}), which has a monitoring program to obtain time delay between multiple images of lensed quasars. WGD2038-4008 follow-up is carried out in MPIA 2.2m telescope (La Silla Observatory, Chile) with an average of one measurement per week (F. Courbin private communication), although no time delay has been measured yet. 
We considered three dates that were 7 days apart and within 2 weeks of our X-shooter observations, then shift the B data to correct by time delay, and estimate the average magnitude difference as  
$(m_A-m_B)_{corr} \sim0.16 \pm 0.03$. This value is in good agreement with our estimation using the core of the emission lines. Therefore, $\Delta m= -0.16 \pm 0.06$ mag seems to indicate the presence of constant/or long-lasting microlensing event not detected by the light curves (\citealt{2014Sluse}). 
To investigate this possibility, we estimate the time scale associated to such event. Following \citealt{2004Treyer}, we define two time scales: 1) the standard lensing time ($t_E$), and 2) the crossing time ($t_{cross}$). The first one represents the time it takes a star to cross a length equivalent to the Einstein radius:
\begin{eqnarray}
t_E = (1+z_L) R_E /v_\perp
\end{eqnarray}
\noindent
where z$_{L}$ is the lens redshift, $R_E$ the Einstein radius in the source plane and $v_\perp$ the effective source velocity (\citealt{2004Treyer}). The second time scale, refers to the time that, within the length of an Einstein radius, the source may encounter a caustic line, causing a large magnification:
\begin{eqnarray}
t_{cross} = (1+z_L) R_{source} / (v_\perp (D_S / D_L))
\end{eqnarray}
\noindent
where $R_{source}$ is the quasar size (i.e. the accretion disc size, $\rm log_{10}(r_{s}/cm$) = 15.65), and $D_{S}$ and $D_{L}$ the angular diameter distance (\citealt{1999Hogg}) of the observer-source and observer-lens respectively with our assumed cosmology. 
Considering a typical value of $v_\perp$ = 600 km s$^-1$ as assumed by \citealt{2004Treyer}, we estimate $t_E \simeq 27.7$ years and $t_{cross}\simeq 0.6$ year.  
On the other hand, if we calculate the effective source velocity following \citealt{2011mosquerakochanek}\footnote{Notice that in \citealt{2011mosquerakochanek} these time scales are defined as $t_{cross} = R_{source}/v_{eff}$ and $t_E= R_E/v_{eff}$, where $v_{eff}$ is the effective velocity and is defined as a combination of the motions of the observer, the lens, and the source.}, we obtain $v_{eff}$ = 820 km s$^-1$, which yields $t_E\sim 20.3$ years and $t_{cross} \simeq 0.5$ year, respectively. Thus,  a long-lasting microlensing event would last for $\sim 20$ years, while the crossing time should be around 6 months. 

\begin{figure}
 \centering
  \includegraphics[width=8.5cm]{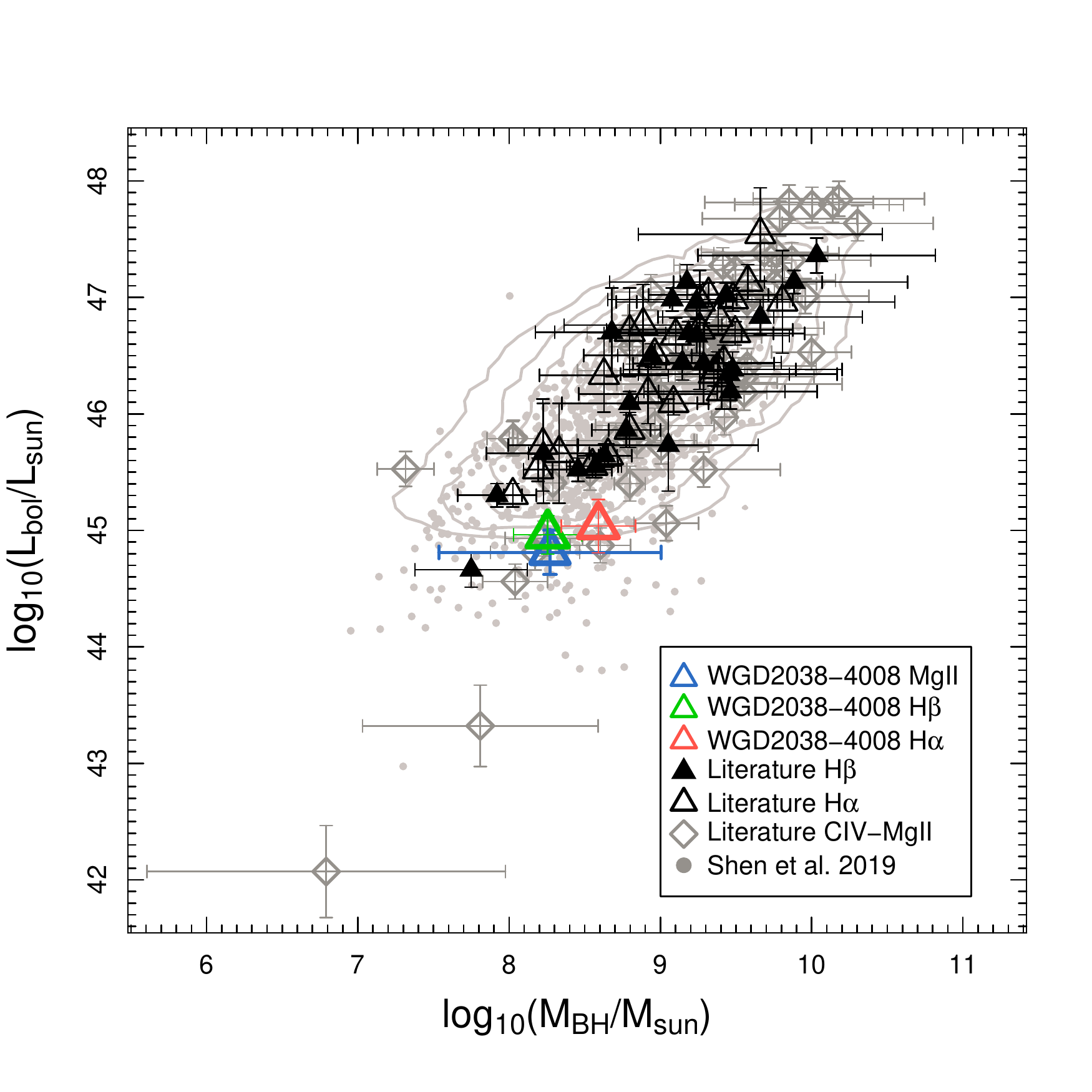}
  \caption{M$_{BH}$ vs L$_{bol}$ for quasars. The masses plotted are estimated from different emission lines and monochromatic luminosity that exists upto date (\citealt{2006Peng,2011assef,2012Sluse,2017ding}). For lensed quasars, black open diamonds correspond to the M$_{BH}$ derived from Mg{\footnotesize{II}} and CIV emission line, black triangle correspond to H$\alpha$ and black filled triangle to H$\beta$ emission line. For non-lensed quasars, we included data from \citealt{2019shen} for M$_{BH}$ from SDSS represented by the grey dots and grey contours. The average $M_{BH}$ mass estimation for WG2038-4008 is represented as blue/red/green triangles for MgII/Hbeta/Halpha emission lines.}
  \label{fig:mbh1}
\end{figure}

In spite of this microlensing magnification, the induced error in the luminosity is negligible as the majority of the error budget is introduced by the macro model magnification (Sec.\ref{sec:Method}).\\
The monochromatic luminosity at L$_{5100}$ was estimated using a single-power law function between two continuum windows on each side of the BELs. It agrees for H$\alpha$ and H$\beta$ of each image, within the errors, with an average value of $\rm log_{10}(L_{5100}/L_\odot$) = 44.29 $\pm$ 0.03. Due to flux loss in UVB, we modeled a SED to estimate L$_{3000}$~ using the magnitudes and magnification of image A, obtaining log$_{10}(L_{3000}/L_\odot$) =  44.23 $\pm$ 0.19. The luminosities L$_{300}$ and L$_{5100}$ agree within their errors, even though they were obtained with different methods.
The M$_{\rm BH}$ was obtained following eq.~\ref{eq:two} with an average value between A and B images of $\rm log_{10}(M_{BH}/M_\odot$) =  8.59 $\pm$ 0.35, 8.25 $\pm$ 0.32, 8.27 $\pm$ 1.06 for H$\alpha$, H$\beta$ and Mg{\footnotesize{II}}, respectively. The M$_{BH}$ estimates obtained using the three different emission lines are consistent within 2 sigma.
We show the M$_{\rm BH}$ estimations along with those of the literature of lensed quasars in Figure~\ref{fig:mbh1}. To avoid the discrepancies associated to the different parameter values used by the authors, we combine their FWHM and monochromatic luminosity values using equation~\ref{eq:two} to obtain M$_{\rm BH}$. We converted from intrinsic to bolometric luminsity applying $\rm L_{bol} = A ~ L_{ref}$, where A = (3.81, 5.15, 9.6) for $\rm L_{ref} = ( L_{1350}, L_{3000}, L_{5100})$ presented in \citealt{2012Sluse}. M$_{\rm BH}$ estimates for 33 lensed quasars are also included in Fig.~\ref{fig:mbh1} (some of them having several values as they are obtained from different emission lines) as well as those of \citealt{2019shen} for (non-lensed) quasars from SDSS reverberation mapping. The Figure shows that our results for image A and B of WGD2038-4008 are in good agreement with those of the non-lensed quasars, situating our object in the low luminosity range of the diagram.\\ 
We can also infer the accretion disk size (r$_s$) assuming a thin-disk model (\citealt{1973shakurasunyaev}) and considering our M$_{BH}$ estimate (\citealt{2011mosquerakochanek}) as:

\begin{equation}
    r_s= 9.7 x 10^{15} \left( \frac{\lambda_{rest}}{\mu m}\right)^{4/3} 
    \left( \frac{M_{BH}}{10^9 M_{\odot}}    \right)^{2/3}  \left( \frac{L}{\eta L_E}   \right)^{1/3} \text{[cm]} \label{eq:ads}
\end{equation}

\noindent $\lambda_{rest}$ is the wavelength where the M$_{BH}$ is measured, $\eta$ is the accretion efficiency and $L/L_E$ the luminosity in units of the Eddington luminosity. For a typical accretion rate $\eta$ = 0.1 and  $L/L_E \sim$ 1/3 (\citealt{2010schulze}). Using the different M$_{\rm BH}$ estimates with the wavelength value at H${\alpha}$, H${\beta}$ and Mg{\footnotesize{II}},  the accretion disk size measurements are shown in Table~\ref{table:tablembh}. The size  Mg{\footnotesize{II}} has an average of $\rm log_{10}(r_{s}/cm$) = 14.98 $\pm$ 0.84, H$\beta$ is 15.25 $\pm$ 0.4 and H$\alpha$ 15.67 $\pm$ 0.74. Our estimates are in agreement within each other and  with \citealt{2018Morgan}. We scaled our wavelength ($\lambda$ in which the M$_{BH}$ was measured) to $2500~\AA$, assuming $r_s \propto \lambda^{4/3}$ and obtained $\rm log_{10}(r_{s}/cm$) = 14.94 $\pm$ 0.22,  15.25 $\pm$ 0.82 and 15.65 $\pm$ 0.79 in Mg{\footnotesize{II}}, H${\beta}$ and H${\alpha}$ respectively. These values are consistent with the theoretical values estimated by \citealt{2018Morgan} at r$_{2500}$: 15.41 $\pm$ 0.15 for Mg{\footnotesize{II}}, 15.37 $\pm$ 0.26 for H${\beta}$ and 15.62 $\pm$ 0.18 for H${\alpha}$.  

\section{Conclusions\label{sec:conclusion}}

We have obtained high S/N observations for the quadruple lensed system WGD2038-4008 using X-shooter instrument in VLT. We use Gaussian fitting to obtain uncontaminated spectra for the A and B lensed quasar images and the lens galaxy. The most prominent emission lines are detected (Mg{\footnotesize{II}}, H$\beta$ and H$\alpha$) as well as absorption lines in the lensing galaxy.\\
We confirmed the velocity dispersion of the lensing galaxy spectra, obtaining $299 \pm 12$ km/s, in agreement with previously estimated values ($2.96 \pm 19$ km/s \citealt{Buckley-Geer2020}).\\
The magnification factors were estimated from the lens parameters of \citealt{shajib2019} ($\mu_A$ = 2.27 $\pm$ 0.21 and $\mu_B$ = 2.71 $\pm$ 0.32) and were used to demagnify the spectra.  Comparing the continuum-subtracted emission lines, we notice there is an enhancement in the right wing of H$\alpha$ of image B, that could be due to microlensing. However, this effect is not seen in H$\beta$ (a region similar in size to H$\alpha$) but this might because of the low S/N and to the presence of absorption-like features.  Mg{\footnotesize{II}} profile does not show any sign of microlensing, and it could be because it is produced in a farther away region. Magnification in the red wing of the H$\alpha$ broad emission line has been detected in HE0435-1223 (\citealt{2014Braibant}) and QSO2237+0305 (\citealt{2016Braibant}). The main conclusion is that these line profile distortions are explained by the differential magnification of a Keplerian disk model. As the continuum region is expected to be smaller than the BLR, those profile distortions are also accompanied by larger magnification of the continuum. However, in our case the magnification in the continuum is smaller than the one in the H$\alpha$ broad emission line. 
On the other hand, several papers describe an  enhancement in the Fe K$\alpha$ profile with higher magnification than the X-ray continuum in  MG J0414+0534 (\citealt{2002Chartas}),  QSO 2237+0305 (\citealt{2003Dai}) and H1413+117 (\citealt{2004Chartas}). This effect is attributed to differential microlensing. \citealt{2003Popovic} use a standard accretion disk and caustic crossing to investigate the structure that could lead to such differences. The authors conclude that different dimensions for the emitting region (e.g. an inner BEL anuli radius smaller than the continuum disk) and `segregation of emitters allow us to reproduce the Fe K$\alpha$ enhancement without equivalent continuum amplification'. Furthermore, \citealt{2007Abajas} demonstrated that this result could also be obtained in the case of a biconic model for the BEL. Thus, a similar effect might be used to explain our results, but a further analysis is needed to confirm this.\\
The FWHM was measured for the three emission lines and are in agreement for H$\beta$ and Mg{\footnotesize{II}} for both images. Even though H$\alpha$ has a discrepancy in the right wing, we measured  the FWHM for both of them (with a difference of $>5$ sigma).\\
Microlensing effect in the continuum was investigated obtaining the magnitude difference for the continuum (0.01 $\pm$ 0.03 mag) and the core of the emission lines (0.17 $\pm$ 0.05 mag). Our values are in agreement with spectroscopic data from \citealt{Nierenberg2020} and with photometric data corrected by time-delay. There seems to be microlensing effect in the continuum of $\Delta m = -0.16 \pm 0.06$ mag.\\
The monochromatic luminosity at 5100~\AA~ was obtained for H$\alpha$ and H$\beta$ using a single-power law function to the region of interest. The luminosities for both images are in good agreement, with a mean of $\rm log_{10}(L_{5100}/L_\odot$) = 44.29 $\pm$ 0.20. On the other hand, L$_{3000}$ was estimated using SED and obtained $\rm log_{10}(L_{3000}/L_\odot$) = 44.23 $\pm$ 0.19. Both luminosities are in agreement within errors.\\
The M$_{BH}$ was measured with the luminosity and the FWHM from the broad emission lines, obtaining a consistent mass for both images in the same BEL and a mean mass of $\rm log_{10}(M_{BH}/M_\odot$) = 8.37 $\pm$ 0.40 for this quadruple lensed quasar.   
When combined with the quasar's monochromatic luminosities, we find Eddington ratios similar to those measured in the literature for unlensed low-luminosity quasars. Finally, we got the accretion disk size from equation~\ref{eq:ads} obtaining an average size of $\rm log_{10}(r_{s}/cm) = 15.28 \pm 0.63$.\\

\begin{table*}
\caption{FWHM, luminosities and M$_{BH}$. }   
\label{table:tablembh}      
\centering   
\begin{tabular}{ l l l l l l c c }     
\hline\hline    
Image & Line  & FWHM [km/s] & log$_{10}$(L$_{ref}$)[L$_\odot$]\tablefootmark{a} 
& log$_{10}$( M$_{BH}$) [M$_\odot$] & log$_{10}$ ( r$_s$) [cm]\tablefootmark{b}  & S/N Line\tablefootmark{c} & S/N Continuum \\
\hline  
  &  Mg{\footnotesize{II}}        &  
3914.52 $\pm$ 500.09 &
 44.23 $\pm$ 0.19 &
 8.25 $\pm$ 0.59 & 
  14.92 $\pm$ 0.31 &
  30 & 5 \\
A &  H$\beta$  &  
4689.32 $\pm$ 42.96 & 
44.29   $\pm$ 0.17 &
8.27 $\pm$ 0.24 &
15.26 $\pm$ 0.79 &
16 & 6 \\
  &  H$\alpha$ &  
5595.68 $\pm$  125.92 &
  44.36   $\pm$ 0.23 &
 8.57 $\pm$ 0.22 &
  15.63 $\pm$ 0.83 &
  73 & 11  \\
  \hline   
  &  Mg{\footnotesize{II}}        &  
4118.73 $\pm$ 921.90 &
  44.23 $\pm$ 0.19 &
8.29 $\pm$ 0.88 &
  14.95 $\pm$ 0.23 &
  35 & 6\\
B &  H$\beta$ &
4817.63 $\pm$ 48.15 & 
44.21   $\pm$ 0.16 &
8.24 $\pm$ 0.21 &
15.23 $\pm$ 0.85 &
19 & 6\\
&  H$\alpha$  &
6150.98 $\pm$ 133.39 &
44.29 $\pm$ 0.23 &
8.61 $\pm$ 0.27 &
15.66 $\pm$ 0.74 &
85 & 12\\
\hline 
\end{tabular}
\tablefoot{
\tablefoottext{a}{L$_{ref}$ = Luminosity ( L$_{3000}$ , L$_{5100}$, L$_{5100}$) for Mg{\footnotesize{II}}, H$\alpha$ and H$\beta$ respectively. The luminosity for H$\alpha$ and H$\beta$ is from the spectra and for Mg{\footnotesize{II}} is obtained from the SED.}
\tablefoottext{b}{r$_s$ is the accretion disk size obtained from equation~\ref{eq:ads} at the $\lambda_{rest}$ of the emission line used for the M$_{BH}$ measurement.}}
\tablefoottext{c}{This is the maximum S/N in the peak of the emission line.}
\end{table*}

\begin{acknowledgements}
A.M. acknowledges grant support from project CONICYT-PFCHA/Doctorado Nacional/2017 folio 21171499. V.M. acknowledges partial support from Centro de Astrofísica de Valparaíso. V.M. acknowledges support from Redes \#190147 (ANID). N. G. acknowledges financial support from ICM Núcleo Milenio de Formación Planetaria, NPF.
N. G. acknowledges grant support from project CONICYT-PFCHA/Doctorado Nacional/2017 folio 21170650. R.J.A. was supported by FONDECYT grant number 1191124.
\end{acknowledgements}

\bibliographystyle{aa} 
\bibliography{aanda} 

\end{document}